\documentclass{emulateapj}
\usepackage{epsf}
\usepackage{apjfonts} 
\usepackage{xspace} 
\usepackage{amsmath}
\def\figdir{.}

\def\eps{\epsilon}

\def\Msun{\dim{M}_{\odot}}

\def\kms{\dim{km/s}}

\def\Luv{L_{\rm UV}}
\def\Mtot{M_{\rm TOT}}

\def\xeff{x_{\rm eff}}

\def\simlt{\mathrel{\rlap{\lower 3pt\hbox{$\sim$}}\raise 2.0pt\hbox{$<$}}}
\def\simgt{\mathrel{\rlap{\lower 3pt\hbox{$\sim$}} \raise 2.0pt\hbox{$>$}}}

\def\ngb{N_{\gamma/a}}
\def\ngbs{N_{\gamma/a, *}}

\def\dim#1{\mbox{\,#1}}
\def\figname#1#2{\figdir/#1}

\def\hide#1{}

\begin{document}

%=================================================
\title{Relative Role of Stars and Quasars in Cosmic Reionization}
%=================================================

\author{Marta Volonteri\altaffilmark{1} \& Nickolay Y.\ Gnedin\altaffilmark{2,3,4}}
\altaffiltext{1}{Department of Astronomy, University of Michigan, 500 Church
Street, Ann Arbor, MI, USA; martav@umich.edu}
\altaffiltext{2}{Particle Astrophysics Center, Fermi National
Accelerator Laboratory, Batavia, IL 60510, USA; gnedin@fnal.gov}
\altaffiltext{3}{Kavli Institute for Cosmological Physics, The
University of Chicago, Chicago, IL 60637, USA}
\altaffiltext{4}{Department of Astronomy \& Astrophysics, The
University of Chicago, Chicago, IL 60637 USA}

\begin{abstract}
We revisit the classical view that quasar contribution to the
reionization of hydrogen is unimportant. When secondary ionization are
taken into account, in many plausible scenarios for the formation and
growth of supermassive black holes quasars contribute substantially or
even dominantly at $z\ga8$, although their contribution generally falls
below that of star-forming galaxies by $z=6$. Theoretical models
that guide the design of the first generation of redshifted 21 cm
experiments must, therefore, substantially account for the quasar
contribution in order to be even qualitatively accurate.
\end{abstract}


%----------------------
\section{Introduction}
\label{sec:intro}
%----------------------

Modeling theoretically the properties of the sources that re-heated
and re-ionized the intergalactic medium at the end of the cosmic Dark
Ages is difficult, because, at present, our knowledge of the nature
of ionizing sources is highly incomplete. The conventional view, held
over a decade since the original work of \citet{Shapiro1987,Shapiro1994, Giroux1996,igm:mhr99}, is that
the ionization at the highest redshifts is dominated by the UV
radiation from star-forming galaxies, and that the non-thermal
contribution from quasars builds up only later, at $z<4$, to dominate
the double ionization of helium (which requires photons of
energies exceeding $54.4\dim{eV}$). 

This conventional view, however, is incomplete. In particular, it
ignores an important physical process of ``secondary ionizations'' by
energetic ionizing photons. It is a well-known physical fact that an
ionizing photon with the energy $E$ in excess of $\sim100\dim{eV}$ is capable
of ionizing more than a single atom - the excess photon energy of
$E-13.6\dim{eV}$ is deposited in the electron, which is capable of
ionizing one or more additional atoms in its vicinity
\citep{atom:ss85,Valdes2008}. 

These secondary ionizations can substantially increase the relative
contribution of quasars to the reionization of the universe. For
example, a $1\dim{keV}$ X-ray photon can ionize up to $\sim25$ hydrogen
atoms. While this important effect has been included in many of the
prior work on hydrogen reionization
\citep{igm:gs96,igm:o01,igm:vgs01,ng:rgs02a,igm:mba03,igm:ro04,igm:dhl04, Madau2004, igm:f06,igm:ztss07,igm:pdc07,Salvaterra2007,Cohn2007,igm:tz08,igm:kh08,igm:sv08,igm:wgw08,igm:sbk08,igm:sapt08,Ripamonti2008}, 
it may be worth re-assessing the canonical view of the sub-dominance of
the quasar population as the source of hydrogen reionization in light
of recent improvements in the values of cosmological parameters and
new developments in our understanding of the growth of supermassive
black holes at high redshifts.

There are two possible approaches to modeling reionization. In the
first one, a model for the emission, propagation, and absorption of
ionizing radiation from various categories of sources is
constructed. Ultimately, such a model must involve a cosmological
numerical simulation that resolves Lyman limit systems with sub-kpc
resolution and clustering of galaxies and quasars on $\sim100\dim{Mpc}$
scales. Such a large uniform dynamic range is not yet feasible in modern
cosmological simulations, so this type of modeling inevitably suffers
from the unknown systematic biases, but, at the same time, it offers
the most complete model of cosmic reionization.

A second, much simpler (and, therefore, much more limited)
approach consists in ignoring the absorption and propagation of
ionizing photons, and restricting a theoretical model to only
counting the number of ionizations per atom. A major (but not the
sole) limitation of this 
approach is that the criterion for reionization - the required number
of ionizations per atom - is unknown. While reasonable
estimates for this number can be constructed \citep[e.g.][and
  discussion there]{ng:g08}, it, at the very least, must exceed 
  unity\footnote{Or, even more precisely, about 0.97, as a few percent of gas
  remains neutral after reionization, locked in Lyman limit and Damped
  Ly$\alpha$ systems.}.

In this brief paper, we adopt the second, simple approach, to couple
theoretical expectations for both stellar sources and quasars, and to
consider their relative contribution to the total number of
ionizations per hydrogen atom. In line with our limited goals, we
restrict our attention only to two main astrophysical types of
ionizing sources, although we remain acutely aware that other, more
exotic sources - including the energetic photons and electrons from
dark matter annihilation - can provide a substantial or even dominant
contribution to the total ionization budget \citep[e.g.][and
  references therein]{igm:bh09}.

%----------------------
\section{Method}
\label{sec:method}
%----------------------

We evolve the population of two distinct types of seed massive black holes (MBHs): either ``small seeds'', derived from
Population III remnants \citep{Madau2001,VHM}, or ``large seeds", derived from gas-dynamical 
collapse in metal--free galaxies \citep{Begelman2006,Lodato2006}. 
We adopt one single formation scenario for each realization of the Universe we consider, 
therefore each model contains only ``small seeds'' or ``large seeds''.
 
In the Population III remnants model, seed MBHs form with masses
$m_{\rm seed}\sim$ few$\times10^2\Msun$ (see Fig.~1, top left panel), in haloes collapsing at
$z>15$ from rare 3.5-$\sigma$ peaks of the primordial density field
\citep{VHM}. 
%We assume that seeds form in the mass range
%$125<M_{\rm BH} <1000\,\Msun$, from an initial stellar mass function
%with slope $-2.8$. 

In the ``large seeds'' scenario, massive seeds with $M\approx
10^4\Msun$ can form at high redshift, when the
intergalactic medium has not been significantly enriched by metals.  
Here we refer to \citet{Begelman2006,Lodato2006}, for more details of the physical 
MBH formation model.  Seeds form in gravitationally unstable 
pre-galactic disk with primordial composition, in halos with virial temperature $\sim 10^4$ K, cooled mainly by atomic hydrogen. 
The stability of gaseous disks depends on two parameters, the halo spin parameter $\lambda_{\rm spin}$, and the 
fraction of baryonic matter that ends up in the disk, $f_d = (\Omega_M/\Omega_b) (M_{\rm disk}/M_h)$.  
%Christoudoulou et al. (1995) propose a simple, but robust,
%criterion for stability which can be expressed as $(\frac{1}{2}f\,T_K/|W|)^{1/2}<0.34$,
%where $T_K$ is the rotational kinetic energy, $W$ is the gravitational potential energy, 
%and $f$ is a parameter dependent on the geometry of the system, with $f=1$ for disks. 
A maximum spin parameter $\lambda_{\rm spin,max}$ exists, for which a disk is unstable as 
a function of the fraction of baryons forming the disk,  i.e., for every $f_d$, disks are stable 
for $\lambda_{\rm spin}>\lambda_{\rm spin,max}$. We here assume ($f_d$, $\lambda_{\rm spin,max}$)=(0.2,0.2).
The mass of the forming MBH seeds
is set by the joint characteristics of the gas flow and of the evolution of the collapsed gas \citep{Begelman2007}.
The mass function of MBH seeds peaks at $10^4\Msun$ (see Fig.~1, top right panel). 

We study MBH evolution within dark matter halos via a Monte-Carlo algorithm based on the
extended Press-Schechter formalism. The population of MBHs evolves along with their hosts according to a ``merger driven scenario'', as
described in Volonteri et al. (2003; 2009). An accretion episode is assumed to occur 
as a consequence of every major merger (mass ratio larger than 1:10) event. 
During an accretion episode, each MBH accretes an amount of mass, 
$\Delta M=9\times 10^7\Msun(\sigma/200\kms)^4$, that scales with the $M_{\rm
BH}-\sigma_*$ relation of its hosts (see Volonteri \& Natarajan 2009). Accretion starts after a
dynamical timescale and lasts until the MBH has accreted $\Delta M$.

We model the accretion rate onto MBHs in two ways. As a baseline model, we assume that accretion proceeds at
the Eddington rate \citep[see also][]{Salvaterra2007}. In a second case, we model the accretion rate 
during the active phase from the extrapolation of the 
empirical distribution of Eddington ratios, $\lambda=\log(L_{\rm
bol}/L_{\rm Edd})$, found in \cite{Merloni08}. We adopt a fitting
function of the Eddington ratio distribution as a function of MBH
mass and redshift (Merloni 2009).  We are here extrapolating such a model at much higher 
redshifts and lower MBH masses than originally intended. We caution readers in taking the results of this model
face value.The main goal of our exercise is to probe possible sensible ranges for the accretion rates on MBHs. 

% {\bf [Since this model is based on an extrapolation in both mass and redshift, I think we should discuss the model and its results, but it's not necessary to show plots.]}
%\note{NG: ok}

%The distributions of Eddington ratios,
%$f_\lambda$, are computed in 10 redshift intervals (from $z=0$ to
%$z=5$) for 4 different mass bins (for the mass range $6 < \log(M_{\rm BH}/\Msun)< 10$), 
%and then fit with an analytic function which is the sum of a Schechter function and a log-normal. 

At each timestep of our Monte Carlo simulation of halos merger trees, we calculate the 
average energy density emitted during each timestep by the MBH population as follows:
%\note{NG: I added deltas as these are changes}
\begin{equation}
\Delta U=\langle \eps \rangle \Delta\rho_{acc}c^2, 
\label{eq:enden}
\end{equation}
where $\Delta\rho_{acc}$ is the total mass density (in comoving units) accreted by MBHs within the timestep,
and $\langle \eps \rangle$ is the average radiative efficiency, which we assume depends solely on MBH spins for radiatively efficient accretors.
We evolve MBH spins according to two simple models: coherent accretion \citep{Volonteri2005} and chaotic accretion \citep{King2007}. These two models lead to rapidly spinning MBHs (``high spins'', $\langle \eps \rangle=0.2$), and slow rotators (``low spins'', $\langle \eps \rangle=0.06$) respectively.  We assume that MBHs are born non-spinning.

\begin{figure}[t]
\plotone{\figname{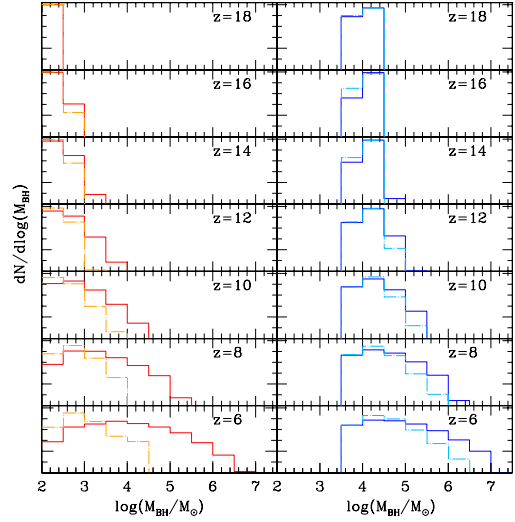}{f0.eps}}
\caption{Mass function of MBHs at different cosmic times. Top to bottom: $18<z<20$; $16<z<18$; $14<z<16$; $12<z<14$; $10<z<12$; $8<z<10$; $6<z<8$. Left panels: small seeds. Right panels: large seeds. Solid histogram: Eddington accretion; dot-dashed histogram: Merloni-Heinz accretion.\newline }
\label{fig:bhmf}
\end{figure}

The energy density, $U$, depends on both the accretion rate (in Eddington units, $f_{Edd}=10^\lambda$) and $\eps$.
Assuming constant $f_{Edd}$ and $\eps$, at a given time $t$: 
\begin{equation}
U(t)\propto \eps\, M_{\rm seed}\, \left( e^{f_{Edd}\frac{1-\eps}{\eps}\frac{t}{\tau}}-1 \right)
%U\approx \eps \times M_{\rm seed} \times (\exp(f_{edd}((1-\eps)/\eps)(t/\tau))-1)
\end{equation}
where $M_{\rm seed}$ is the average mass of MBH seeds in each scenario, and $\tau=0.45$ Gyr is the Eddington timescale. 
We can see that if $t\ll\tau$: $U\propto f_{Edd}(1-\eps)$. Within our evolutionary scheme $\eps$ is determined for each MBH, while we have used a single, average, $\langle \eps \rangle$ in calculating $U$. Since it takes a few Myr for MBHs in the coherent accretion model to spin-up to large spins (i.e. for the radiative efficiency to increase from 0.06 -- Schwarzschild hole, to 0.2 -- spin$=0.9$) this scheme slightly overestimates the radiative output from MBHs in the `high spin' case at $z>11$. 

% so that in reality I should define the average eps at a given cosmic time and: $U\propto \eps\times f_{edd}(1-\eps(t))$

We further assume that a fixed fraction $f_{\rm UV}$ of the bolometric power radiated by high-redshift quasars is  emitted as hydrogen-ionizing photons.
%For a given bolometric emissivity,
The number of ionizing photons scales as $f_{\rm UV}/E_\gamma$, where $E_\gamma$ is the {\it mean}
photon energy (see section 2).  We refer the reader to \cite{Madau2004} for a thorough discussion about the quasar spectra;
in the following we assume, conservatively, that  $f_{\rm UV}=0.2$.  

The total comoving energy density of ionizing radiation emitted by the growing MBH population is then:
%\note{NG: I do not undernstand this equation, a physical quantity cannot be integrated over $dz$.}
\begin{equation}
U_\gamma(z)=\sum_{i=0,j(z)} f_{\rm UV}\,\Delta U_i
%U_\gamma(z)=\int_{z}^{z_{\rm max}}  \eps_{UV}\,U\, dz,
\label{eq:ngamma}
\end{equation}
where the sum is over the timesteps between our starting redshift ($z_{\rm max}=20$, timestep 0), and the timestep, $j(z)$ corresponding to the redshift of interest, $z$. 

We have also implemented a more pessimistic case for MBH accretion, loosely based on \cite{Milosavljevic2008}, 
where the accretion rate has been fixed to 30\% of the Eddington
accretion rate. In this scenario, the yield in ionizing radiation is
lower than for the case of 100\% Eddington rate, but ionization
histories in this scenario fall within the range spanned by our other
models; they, thus, do not change any of our conclusions and we do not
consider them further in this paper.

Our models are consistent with the constraints from the soft X--ray background \citep{Dijkstra2004}. Our predicted population of high redshift AGN  
would account for almost 5\% of the measured (0.5--8 keV) background, or $\sim 25$\% of the unresolved one \citep{Salvaterra2005,Salvaterra2007}. Models are also consistent  with the bolometric luminosity function of quasars at $z>4$ \citep{Hopkins2007}. ``Large seeds'' are preferred as our fiducial model, based on a better agreement with the bolometric luminosity function at $z \sim 4$, however we note that in this paper we probe much higher redshifts than those probed by the study of \cite{Hopkins2007}, hence we consider the match with the luminosity function at lower redshifts as a weak constraint.

%accretion modes:\\
%1) Eddington modes: fixed eddington accretion rate\\
%2) MM models: fixed 30\% eddington accretion rate. Loosely based on  Milosavljevic's model for seed BH accretion. \\
%I implemented a simplified verion, using the average accretion rate suggested  by his simulations rather than the complete model
%3) MH models: the accretion rate is extracted (Monte Carlo) from the  distribution function (DF) of accretion rates (parameters: BH mass, redshift) derived by Merloni \& Heinz using synthesis models. \\

%spin evolution: \\
%1) coherent accretion, which implies large spins\\
%2) chaotic accretion, which implies small spins\\

%The lowest yield in photons is the case MM, high spins. In general models MM give the lowest number of ionizing photons. 

For the stellar contribution to the total ionizing background we use
the extrapolation of the observed UV luminosity functions of high
redshifts galaxies \citep{hizgal:biff07,hizgal:biff08} and the assumed
value for the relative (to the amount of escaped UV light at
$1000\AA$) escape fraction of ionizing radiation. The complete details
of the methodology are described in \citet{ng:g08}. Here we only
briefly repeat that the total mass-to-light ratios of higher redshift
galaxies are computed in a given cosmology by matching the observed
spatial abundance of galaxies of a given luminosity to the
theoretically computed abundance of halos of a given mass, so that
\begin{equation}
  n(>\Luv) = n(>\Mtot)
  \label{eq:nlm}
\end{equation}
(and, optionally, an additional factor that accounts for ``bursty''
star formation rate in high redshift galaxies can be incorporated in
eq.\ [\ref{eq:nlm}]; reasonable values for that factor make
insignificant impact on our results).

The obtained thus mass-to-light ratio can be extrapolated to higher
redshifts. Since the mass-to-light ratio is a weak function of
redshift, the uncertainty of such an extrapolation does not dominate
the final uncertainty of our estimate for the ionizing emissivity from
galaxies. Instead, the final uncertainty is dominated by the
uncertainty (both observational and theoretical) on the adopted value
of the relative escape fraction for ionizing radiation \citep{ng:g08}.

The combined (stellar plus quasar) contribution to reionization can
then be estimated as
\begin{equation}
  \ngb = \ngbs + \frac{U_\gamma}{E_\gamma n_a} + f_{\rm
  SI}(x)\frac{U_\gamma}{14.4\dim{eV} n_a},
  \label{eq:ngb}
\end{equation}
where $\ngb$ is the number of ionizations per atom, 
$n_a=(1-0.75Y_p)n_b\approx2.0\times10^{-7}\dim{cm}^{-3}$ is the
comoving number density of atoms of hydrogen or helium (we assume that
most of helium is only singly ionized during hydrogen ionization), and
$14.4\dim{eV}$ is the mean ionization energy per atom. The first term
in this equation accounts for the contribution from stars, the second
one includes primary ionizations by ionizing photons from quasars with
the mean 
photon energy $E_\gamma$, and the last term accounts for secondary
ionizations from energetic ionizing photons. For the gas uniformly
ionized to the ionization fraction $x$, the fraction of radiation
energy density $f_{\rm SI}$ going into secondary ionizations has been
computed by \citet{atom:ss85}, and their results can be conveniently
fitted by a simple but accurate formula \citep{ng:rgs02a}
\begin{equation}
  f_{\rm SI}(x)\approx0.35(1-x^{0.4})^{1.8} - 1.77\left(\frac{28\dim{eV}}{E_\gamma}\right)^{0.4}x^{0.2}(1-x^{0.4})^2.
  \label{eq:fsi}
\end{equation}
In reality, of course, the universe is not uniformly ionized, so the
quantity $x$ in equation (\ref{eq:ngb}) is an effective value $\xeff$,
such that
\[
  \ngb(\xeff) = \langle \ngb(x) \rangle
\]
and the average is mass-weighted. Thus, $\xeff$ is \emph{not} a mass-
or volume-weighted average of the cosmic ionization 
fraction. However, $\xeff=0$ for the fully neutral and $\xeff=1$ for
the fully ionized universe.

The quantity $\ngbs$ has been computed in \citet{ng:g08} and is not
discussed here. The mean energy of ionizing photons from quasars,
$E_\gamma$, remains a parameter of our model. The 
exact value of this parameter obviously depends on the spectral shape
of the quasar energy distribution. If we assume a classic 
multicolor disk spectrum up to $kT_{\rm max}\sim 1\,{\rm keV}\,(M_{\rm BH}/\Msun)^{-1/4}$ 
(Shakura \& Sunyaev 1973), and a nonthermal power-law component with spectral slope 
$L_\nu\propto \nu^{-\alpha}$, with $\alpha\approx 1$ at higher energies, we find 
%$E_\gamma \simeq 390\dim{eV}$ for $M_{\rm BH}=10^2\, \Msun$, 
%E_\gamma \simeq 300\dim{eV}$ for $M_{\rm BH}=10^3\, \Msun$, and
%$E_\gamma \simeq 180\dim{eV}$ for $M_{\rm BH}=10^4\, \Msun$.
$E_\gamma \simeq 300\dim{eV}$ for $M_{\rm BH}=10^3\, \Msun$.  
The spectrum is harder/softer for smaller/larger MBH masses. 
As a fiducial value, we adopt $E_\gamma=300\dim{eV}$ (cfr. Fig~1), and we 
investigate the sensitivity of our results to the precise value of this 
parameter in the next section.

Equation (\ref{eq:ngb}) cannot be evaluated without an equation for
$\xeff$ as a function of time. A complete computation of $\xeff(t)$ requires a
sophisticated three-dimensional modeling of the transfer of ionizing
radiation throughout the inhomogeneous gas density distribution in the
universe. Since such modeling is well beyond the scope of this paper,
we instead introduce the following ansatz for $\xeff(t)$:
\begin{equation}
  \xeff = \min(1,f_{n\rightarrow x} \ngb),
  \label{eq:x}
\end{equation}
where $f_{n\rightarrow x}$ is a parameter. The motivation of this
ansatz follows from the realization that if every atom in the universe
is ionized exactly once, then in the beginning of reionization, while
$\xeff$ is sufficiently small, $\xeff\approx\ngb$.\footnote{If we
  ignore the second term in equation \ref{eq:fsi}, which is small for
  $E_\gamma\gg100\dim{eV}$, and expand $f_{\rm SI}$ in Taylor series
  around $x=0$, then for sufficiently small $x$, $\xeff=\langle
  x\rangle$.} The factor $f_{n\rightarrow 
  x}$ therefore accounts for the loss of ionizing photons to
recombinations and for more complicated dependence of $\xeff$ on $\langle
  x\rangle$ at the later stages of reionization. This factor cannot be
  too small (or the universe would  
never be reionized), and any value for
$f_{n\rightarrow x}$ above about 0.5 (half of photons lost for
recombination) does not affect our conclusions. Therefore, in the rest
of this paper we adopt $f_{n\rightarrow x}=0.75$ as our fiducial
value. In reality, $f_{n\rightarrow x}$ must be a function of time,
but since it is not likely to be much lower than 1, the exact time
dependence of $f_{n\rightarrow x}$ is unimportant at the level of
precision of our approximations (which is dominated by the
uncertainties on the escape fraction from galaxies and the lack of
knowledge for the specific parameters of our quasar model).

%---------------------
\section{Results}
\label{sec:results}
%---------------------

\begin{deluxetable}{lcc}
\tablecaption{Cosmological Parameters\label{tab:cosmo}}
\tablehead{
\colhead{Parameter} & \colhead{WMAP-5} & \colhead{WMAP-5UP}
}
\startdata\\
$\Omega_m$ & 0.28 & 0.28 \\
$h_{100}$  & 0.7  & 0.7  \\
$\sigma_8$ & 0.82 & 0.84 \\
$n_S$      & 0.96 & 0.98 \\
\enddata
\end{deluxetable}

In the rest of this paper we consider two different cosmological
models. The first one (which we call ``WMAP-5'') is the most likely
joint fit to the fifth year WMAP data, Baryonic Acoustic Oscillations
from SDSS, and supernovae, as presented in the second column of Table
1 from \citet{cosmo:kdnb09}. The second model (``WMAP-5UP'') is
obtained from the first one by increasing both the amplitude of the
density fluctuations $\sigma_8$ and the scalar spectral index $n_S$
upward by 1 standard deviation. The relevant values of cosmological
parameters for these two models are listed in Table \ref{tab:cosmo}.

\begin{figure}[t]
\plotone{\figname{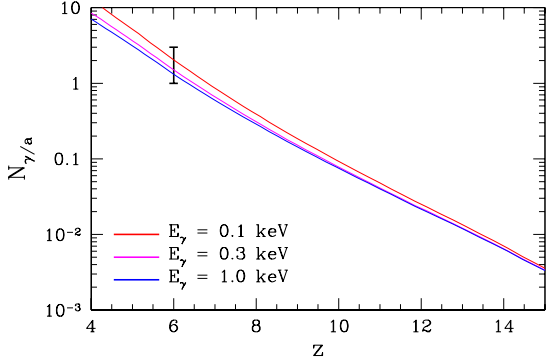}{f0 .eps}}
\caption{The total ionizations-to-atom ratio as a function
  of redshift for our fiducial model (``large seeds, high spins'' in
  WMAP-5UP cosmology) for 3 values of the mean energy of ionizing
  photons from quasars $E_\gamma$. The adopted reionization criterion,
  $1<N_{\gamma/a}<3$ at $z=6$, is shown as a thick black segment with
  error-bars.} 
\label{fig:ngb0}
\end{figure}

Figure \ref{fig:ngb0} demonstrates the sensitivity of the total
ionizations-to-atom ratio $\ngb$ on the assumed value for the mean
energy of ionizing photons from quasars $E_\gamma$. Because
ionizations from the quasar population are dominated by secondary
ionizations (which are independent of $E_\gamma$) for any plausible
value of $E_\gamma$, this parameter has only a mild effect on our results. 

\begin{figure}[t]
\plotone{\figname{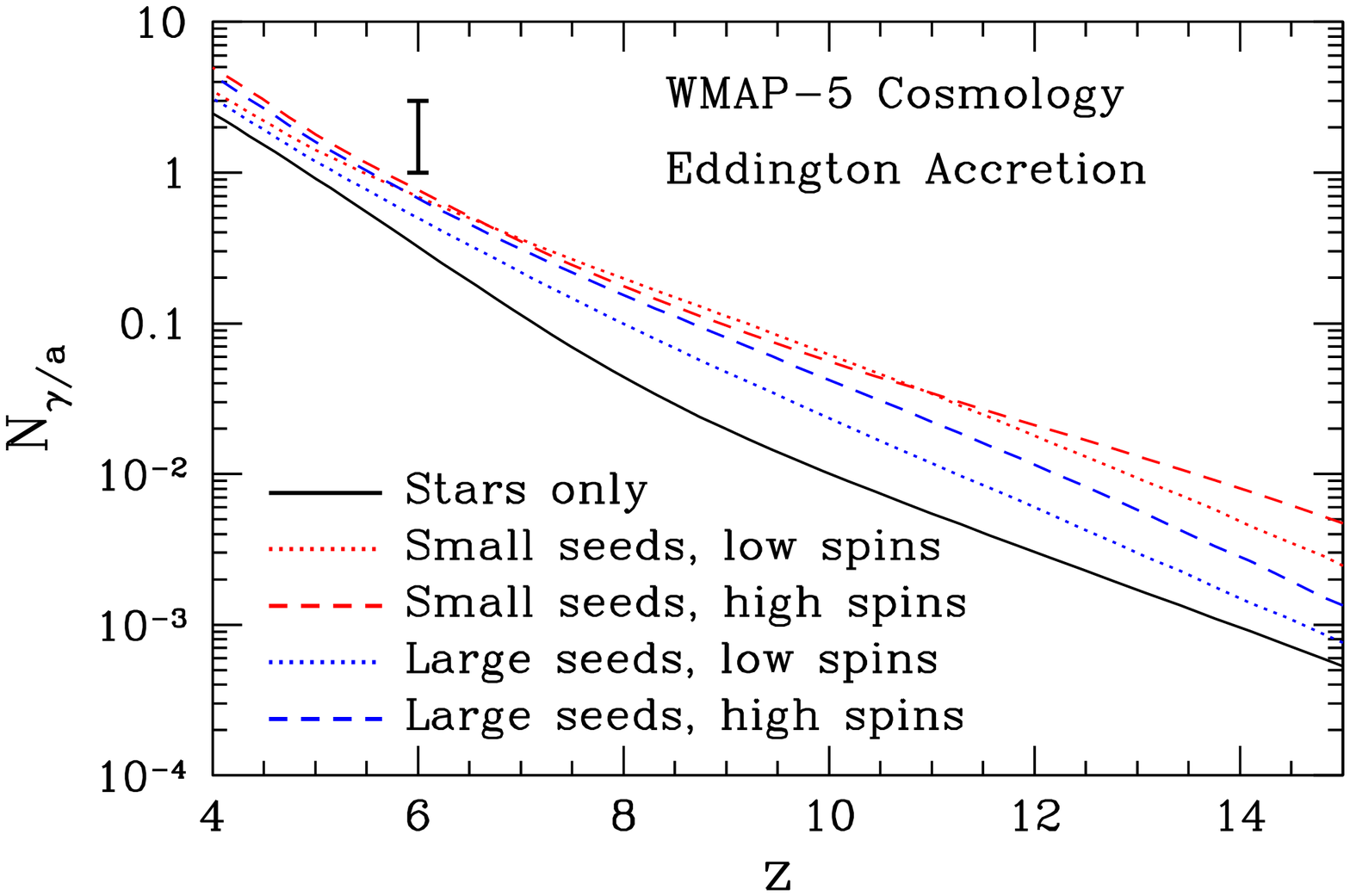}{f1a.eps}}
\plotone{\figname{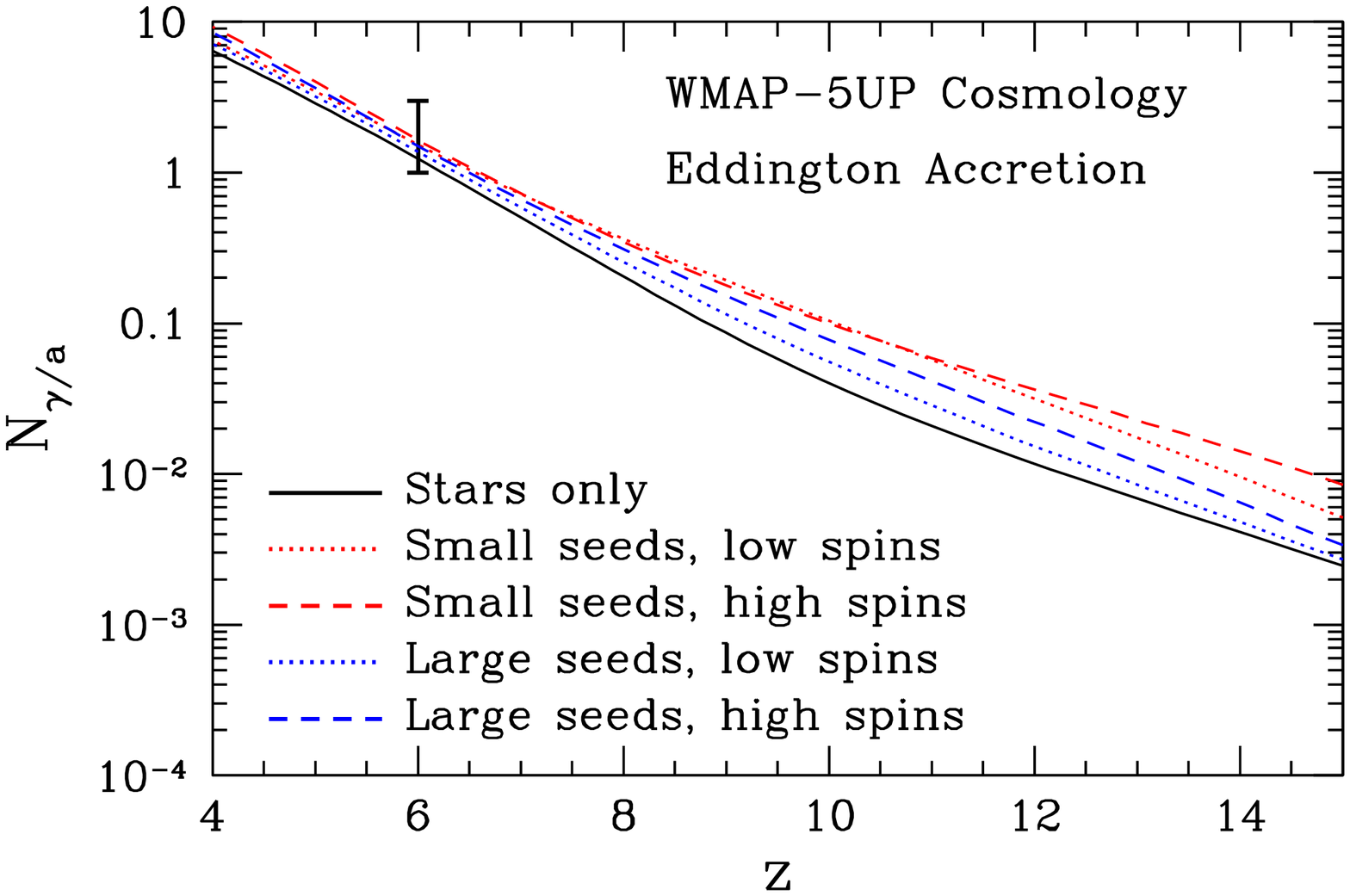}{f1b.eps}}
\caption{The total ionizations-to-atom ratio as a function
  of redshift for different models for quasar contribution (red and blue
  lines) as well as contribution from the stars alone (black solid
  line) for the Eddington accretion model for two chosen
  cosmologies. The adopted reionization criterion, $1<N_{\gamma/a}<3$
  at $z=6$, is shown as a thick black segment with error-bars.}
\label{fig:ngb1a}
\end{figure}

\begin{figure}[t]
\plotone{\figname{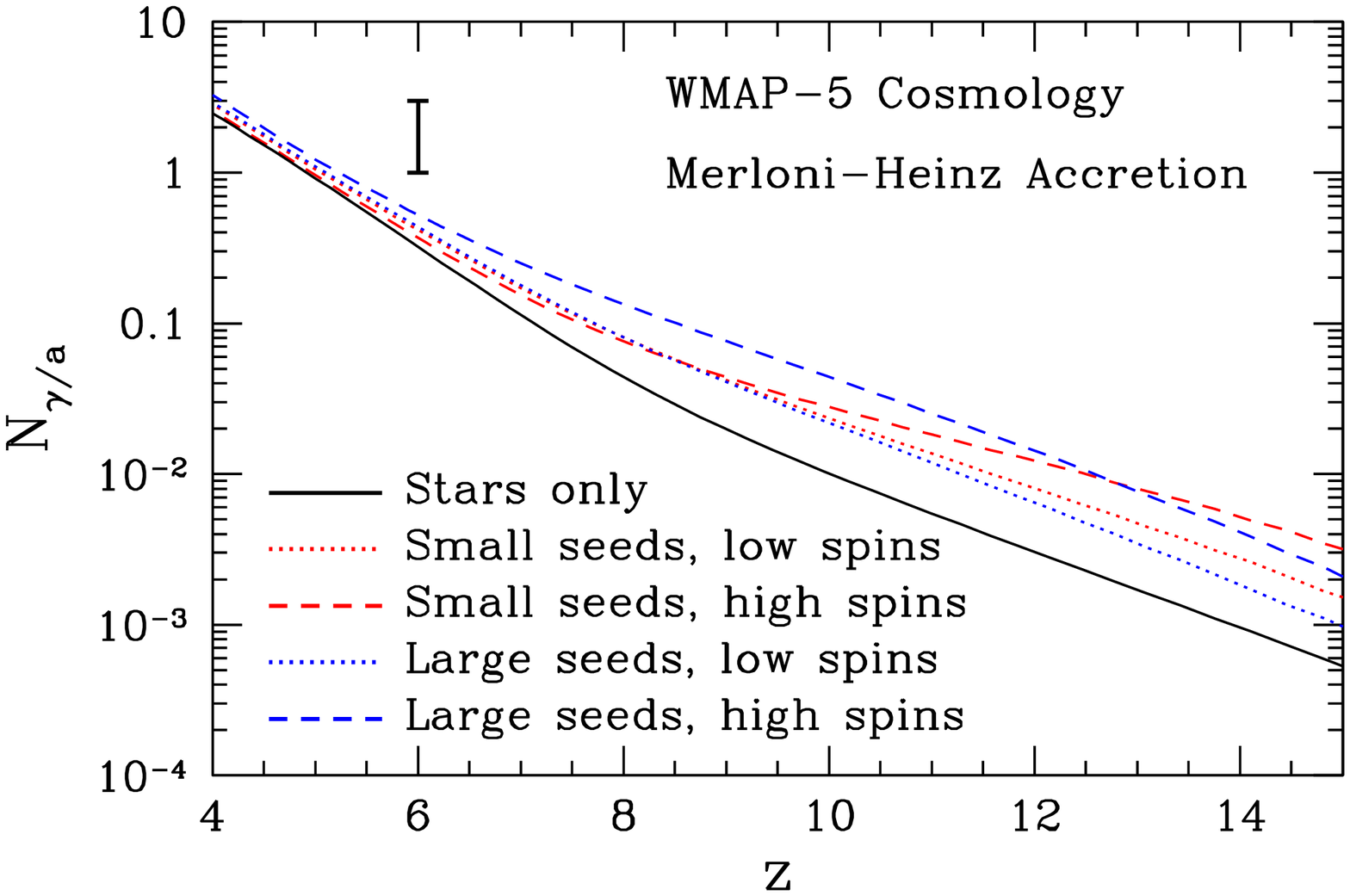}{f2a.eps}}
\plotone{\figname{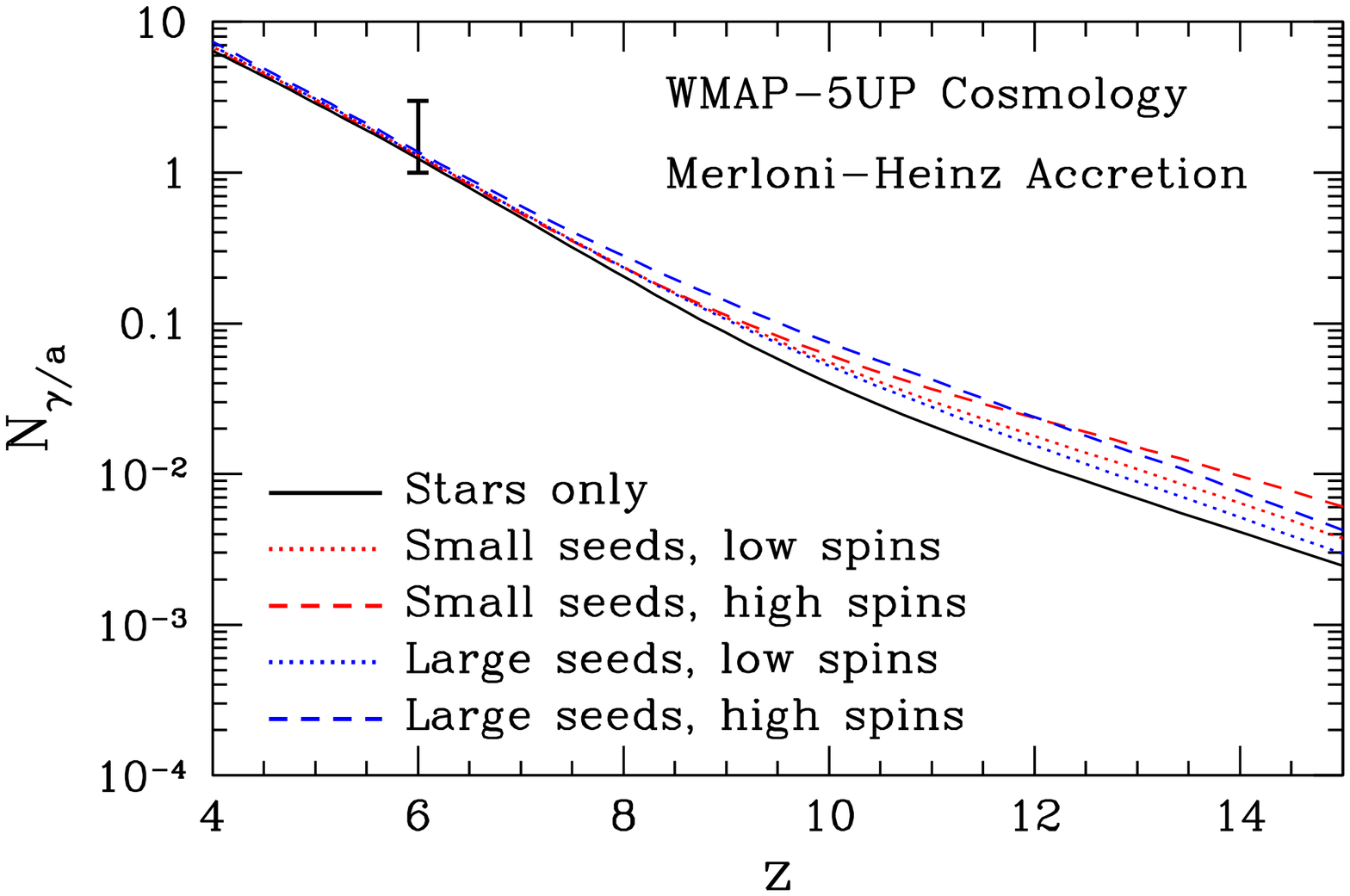}{f2b.eps}}
\caption{Same as Fig.\ \ref{fig:ngb1b}, but for the Merloni \& Heinz
  accretion model.}
\label{fig:ngb1b}
\end{figure}

Figures \ref{fig:ngb1a} and \ref{fig:ngb1b} now present our main
result: the computed $\ngb$ as a function of redshift for two
cosmologies and various parameters of the quasar model. The 
criterion for the universe to be reionized at $z=6$ ($1<\ngb<3$) we
adopt from \citet{ng:g08}, where it is discussed and justified.

\begin{figure}[t]
\plotone{\figname{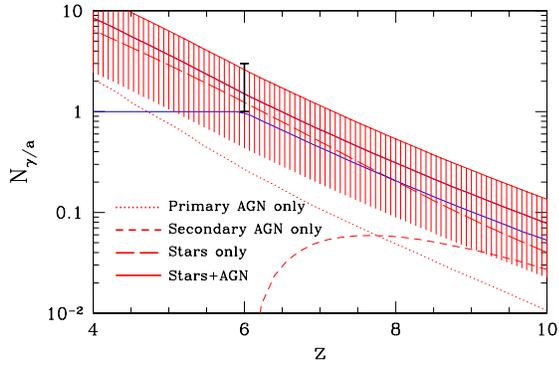}{f3.eps}}
\caption{The total ionizations-to-atom ratio as a function
  of redshift our fiducial model (``large seeds, high spins'' in
  WMAP-5UP cosmology - red lines). The hatched area shows our estimate
  of the observational uncertainty (for a given theoretical
  model). The blue line shows the effective ionization fraction
  $\xeff$ from equation (\ref{eq:x}). Dotted, short-dashed, and
  long-dashed red lines show the contributions from primary and
  secondary ionization from quasars and from stars respectively.} 
\label{fig:ngb2}
\end{figure}

Clearly, the WMAP5 model (with the most likely values of $\sigma_8$
and $n_S$) is somewhat short of the reionization requirement. As has
been discussed by \citet{ng:g08}, this is not necessarily a serious
problem, since the likely uncertainty on our extremely simple model is
a factor of 2 or so. Never-the-less, slightly higher values for
$\sigma_8$ and/or $n_S$ provide a wider breathing space for
reionization modeling. 

In order to illustrate a plausible estimate for the uncertainties due
to observational errors on the escape fraction \citep[c.f.][]{igm:sspa06},
high redshift luminosity functions \citep{hizgal:biff07},  
etc, we show in Figure \ref{fig:ngb2} the ``large seeds - high spin''
quasar model in the WMAP-5UP cosmology (which we adopt as our
``fiducial model'') together with the estimated errors. Notice, that
formally \citet{hizgal:biff07} galaxy luminosity function at $z>6$ should be
considered as an upper limit; we emphasize this by having the hatched
area in Fig.\ \ref{fig:ngb2} unclosed from below. We also show the
evolution of the effective ionization fraction $\xeff$ as 
an illustrative (but highly approximate) reionization history.

%-----------------------------------
\section{Discussion}
\label{sec:discussion}
%-----------------------------------

Since our quasar models cover a wide range of physically plausible
possibilities, we can draw some general conclusions from Fig.\
\ref{fig:ngb1a} and \ref{fig:ngb1b}. In particular, we notice that
(i) by $z\approx6$ the quasar contribution to the total ionization
budget becomes mostly sub-dominant, not exceeding 50\% in the best
case, and likely falling below 20\% or so. Never-the-less, at
$z\approx8$ in all our models quasars contribute from over 50\% up to
90\% of ionizations. This conclusion is particularly important to
theoretical modeling and future observational measurements of the
expected redshifted $21\dim{cm}$ emission from neutral hydrogen in the
reionization era. While most of the
currently planned pathfinder experiments rely, in part, on the
existing simulations for critical design decisions, the predictions
from those simulations - none of which include a quasar contribution -
become inaccurate at $z\ga8$. It is, therefore, important to keep in
mind that much theoretical work still needs to be done before even the
basic observables for the incoming $21\dim{cm}$ experiments (like the
fluctuation power spectrum) can be computed theoretically with 
$\sim$20\% precision (within a given cosmological model).

Most of the contribution to the quasar ionizing budget  is dominated
by MBHs  with mass $<10^6\Msun$. Such small,  low-luminosity MBHs do not contribute to the bright end of the luminosity function of 
quasars, and are therefore difficult to account from simple extrapolations of the luminosity 
function of quasars. These small holes are not hosted in extremely massive galaxies residing 
in the highest density peaks (5 to 6$\sigma$ peaks), but are instead found 
in more common,  ``normal'' systems, $\sim3\sigma$, peaks. 
Future generation of space--based telescopes, such as {\it JWST} and {\it IXO}, are likely to detect and constrain the evolution of the
population of accreting massive black holes at early times ($z\simlt10$).

\acknowledgements 

We are grateful to the anonymous referee for constructive and
ensightful comment and to Andrey Kravtsov for the permission to use
his halo mass function code free of charge.
This work was supported in part by the DOE, by the NSF grant
AST-0507596, and by the Kavli Institute for Cosmological Physics at
the University of Chicago (NG). Support for this work was also provided by 
NASA through Chandra Award Number TM9-0006X (MV).

%\bibliographystyle{apj}
%\bibliography{ng-bibs/self,ng-bibs/atom,ng-bibs/hizgal,ng-bibs/cosmo,ng-bibs/igm,marta}

%\end{document}

\end{document}